\documentstyle[12pt]{article}

\title{Spectral Flow and Vortex Dynamics in Superfluids, Superconductors  and
Ferromagnets }
\author{G.E. Volovik\\
Low Temperature Laboratory, Helsinki University of
Technology\\ FIN-02150 Espoo, Finland, \\  L.D. Landau Institute for Theoretical
Physics\\   Kosygin Str. 2, 117940 Moscow, Russia.}
\begin{document}
\maketitle
\begin{abstract}{
We discuss the nondissipative Magnus-type force acting on linear defects in
Fermi systems, such as Abrikosov vortices in superconductors, singular and
continuous vortices in superfluid phases of $^3$He, magnetic vortices and
skyrmions in ferromagnets.  Spectral flow of fermion  zero modes in the
vortex core gives an essential contribution to the nondissipative force, which
is confirmed in Manchester experiments on rotating $^3$He-B. The spectral flow
effect is closely related to the angular momentum paradox.
}
\end{abstract}

\vskip 0.5 truecm

    Contributed paper to the  XXI International Conference on Low Temperature
Physics, August 8-14, 1996.

\vfill \eject

Let us consider an example which allows us to discuss simultaneously the
vortices in superconductors and  Fermi superfluids, and  magnetic vortices and
skyrmions in ferromagnets. This is quantized vortex line  with continuous
structure, 3 types of such vortices have been experimentally investigated in
superfluid $^3$He-A \cite{AphDiag}. Such vortices can arise in a wide class of
systems, where superfluid (superconducting) correlations are accompanied by
spontaneous orbital or spin momentum. The system is characterized by
fractional number $\ell$ which shows the nominal value of the angular momentum
per particle. For $^3$He-A the momentum is $\hbar$ per Cooper pair, ie
$\ell=1/2$,  while for QHE $\ell=N/2\nu$, where $\nu $ is the filling
factor and $N$ is the number of particles \cite{Read}. Due fermionic degrees
of freedom the  dynamical angular momentum can deviate from its
nominal value ${\bf L}_{nom}= \hbar \ell N\hat {\bf l}$, where $\hat {\bf l}$ is
the direction of spontaneous momentum.  This is the so called {\it angular
momentum paradox}. In $^3$He-A  ${\bf L}_{\rm dyn}=\hbar \ell (N-VC_0)\hat {\bf
l}$, where  $C_0=p_F^3/3\pi^2$ and $V$ the volume \cite{Exotic}. The difference
between particle density $\rho=N/V$ and $C_0$ comes from the particle-hole
asymmetry near the Fermi-surface: in systems with weak interaction it is small
but in strongly correlated systems it can be large.

{\it Continuous vortex}. We consider the following $\hat {\bf l}$-texture
without
singularity
$$
\hat {\bf l}({\bf r}) = \hat {\bf z} \cos\eta(r) +\hat {\bf r}
\sin\eta(r)~~,\eqno(1)
$$
where  $r,z,\phi$ are cylindrical coordinates and $\eta(0)$ is $0$ or $\pi$. The
$\hat {\bf l}$-vector in this texture covers the area $s=4\pi (\cos\eta(\infty)-
\cos\eta(0))$ on unit sphere $\hat {\bf l}\cdot\hat {\bf l}=1$. We are
interested in the textures with $s=2\pi n$, where $n$ is integer. If
$n\neq 0$ there is a circulation of superfluid velocity
${\bf v}_s$ around the texture at infinity:
$\kappa=m\oint d{\bf r}\cdot{\bf v}_s =\hbar s \ell $, as follows from the
Mermin-Ho relation
\cite{Mermin-Ho} ($m$ is the bare mass of the fermion). So the texture
represents the continuous vortex with circulation $\kappa=2\pi \hbar \ell n$.
In $^3$He-A these are Mermin-Ho ($n=1$) and Anderson-Toulouse ($n=2$) vortices.

There are three nondissipative forces acting on the vortex
\cite{KopninVolovik}, which correspond to three velocities involved: the
velocity ${\bf v}_L$ of the vortex;  superfluid velocity
${\bf v}_s$ far outside the vortex; and the
velocity ${\bf v}_n$ of the heat bath (in superconductors and ferromagnets this
is the velocity of the crystal, in superfluids it is the velocity of the
normal component).

{\it Magnus force} ${\bf F}_{\rm M}=\kappa \rho
\hat {\bf z}\times ({\bf v}_L-{\bf v}_s)$ is the force between the
vortex and the superfluid vacuum.

{\it Iordanskii force} ${\bf
F}_{\rm I}=\kappa \rho_n \hat {\bf z}\times ({\bf v}_s-{\bf v}_n)$
arises due to the Aharonov-Bohm effect experienced by the quasiparticles
\cite{Sonin1975}, where $\rho_n$ is the density of the normal
component of the system.

{\it Spectral-flow force} comes
from the spectral flow of fermions through the gap nodes. As distinct from the
most of superfluid/superconductor systems, where the gap nodes appear only in
the core of  singular vortices, in $^3$He-A the point gap nodes exist everywhere
in the bulk liquid. Due to nodes the effect of axial anomaly takes place,
resulting in the transfer of the linear momentum from coherent degrees of
freedom (texture) through the nodes to the heat bath. The momentum flow
occurs with the rate \cite{Exotic}:
$$
\partial_t  {\bf P}=
     {\ell\over {\pi^2}}\int d^2r~ p_F \hat {\bf l} ~(\partial_t {\bf A}
            \cdot (\vec  \nabla \times {\bf A}  \, \, ))  ~~.
\eqno(2)
$$
Here ${\bf A}=p_F \hat {\bf l}$ simulates the vector-potential, which
acts on quasiparticles in vicinity of gap nodes.  Integration of Eq.(2) over
the cross-section of the moving vortex\cite{HydrodynamicAction} gives an
anomalous {\it spectral-flow force}:
$${\bf F}_{\rm
sf}=\kappa C_0 \hat {\bf z}\times ({\bf v}_n-{\bf v}_L) ~~.
\eqno(3)$$
Here $C_0$ is the same parameter as in the {\it angular
momentum paradox}. The
balance of the three forces together with
friction force ${\bf F}_{\rm f}=- D ({\bf v}_L-{\bf v}_n)$ governs the
low-frequency dynamics of continuous vortex.

{\it Singular vortices}. The singular vortex in $s$-wave superconductor or in
$p$-wave $^3$He-B can be obtained from the continuous one in the following way.
Let us take an equal mixture of two states with  $\ell=1/2$ and  $\ell=-1/2$,
then the net momentum in bulk superfluid/superconductor is zero and the gap
nodes
cancel each other. Now, if in each state there is a continuous vortex with
the same $s\ell =\pi n$, this corresponds to the vortex in the mixture with
winding number $n$. The core of this vortex is the region where
the two states do not cancel each other, because they have
opposite $s$, and where the gap nodes survive. By shrinking this region to the
coherence length size one obtains the singular vortex. This procedure just
reflects the core structure of the
$^3$He-B vortex \cite{SalomaaVolovik1987}.

Since the quantity $s\ell =\pi n$ remains the same in this procedure, one may
conclude that all three nondissipative forces remain intact even for singular
vortex. The situation is however more complicated. The anomaly equation (2)
assumes the continuous approximation, which is valid if the discreteness of the
fermionic levels in the core \cite{Caroli} is neglected. The condition for this
is $\omega_0\tau\ll 1$, where
$\omega_0$ is the interlevel distance and $\tau$ is the life-time
\cite{KopninVolovik}. This is the so called "hydrodynamical" regime. So, for
vortices in dirty superconductors, which like the  continuous vortices in
$^3$He-A are in the hydrodynamical regime, the spectral flow is so strong that
it nearly cancels the Magnus force. This means that the Hall conductivity, which
is proportional to $\rho-C_0$, is nonzero only due to small particle-hole
asymmetry \cite{KopninVolovik,vanOtterlo}. Possibly the same approximate
particle-hole symmetry \cite{MakhlinVolovik} leads to cancellation of Magnus
force observed in Josephson junction arrays
\cite{vanZant1995}.

If $\omega_0\tau$ is not small the spectral flow of fermions in the core is
suppressed,  the force ${\bf F}_{\rm sf}$ is modified and finally disappears in
the opposite "collisionless" limit $\omega_0\tau \gg 1$. The parameter
$\omega_0\tau$ is a strong function of the temperature $T$,  this leads to the
$T$-dependence of the spectral-flow force\cite{KopninVolovik}, which reproduces
the experimental $T$-dependence of the nondissipative force acting on the
vortex in $^3$He-B \cite{Bevan1995}. The microscopic derivation of this
temperature dependence has been done many years ago using
the Green's function formalism \cite{KopninCoAuthors}.

{\it Magnetic vortices and skyrmions in ferromagnets}.  Interest to these
defects is renewed because they can represent a new type of  elementary
excitations in QHE (see eg \cite{Stone,ReadSachdev}).
This can be either the cylindrical domain in the easy axis ferromagnet or
$2\pi$ vortex in the easy plane ferromagnet. Such defects can be obtained from
the continuous vortex in Eq.(1) by cancellation of the
superfluid/superconductor long-range order. If superfluidity is absent, one has
$\rho_n=\rho$ and the superfluid velocity drops out from the total
nondissipative force:
${\bf F}={\bf F}_{\rm M}+{\bf F}_{\rm I}+{\bf F}_{\rm sf}= \kappa (\rho -C_0)
\hat {\bf z}\times ({\bf v}_L-{\bf v}_n)$ or
$${\bf F}=s L_{\rm dyn} \hat {\bf
z}\times ({\bf v}_L-{\bf v}_n) ~~.
\eqno(4)$$
Thus the nondissipative force on the
magnetic vortex ($s=2\pi$) and on the skyrmion ($s=4\pi$), is completely
determined by the angular momentum density $L$ and by the area
$s$ covered by the magnetization  on unit sphere. This is in agreement with the
analysis in \cite{SoninNikiforov,FerroWessZumino,Stone,Kuratsuji}. However
due to
the anomalies in the fermionic system, the effective $L$ is the
dynamical one, $L_{\rm dyn}= \hbar \ell (\rho-C_0)$. One may expect that this
corresponds to the "hydrodynamical" regime, while in  "collisionless" regime
the  nominal value of the angular momentum $L_{\rm nom}= \hbar \ell \rho$ is to
be restored.

This work was supported through the ROTA co-operation plan of the Finnish
Academy and the Russian Academy of Sciences and by the Russian Foundation for
Fundamental Sciences.
%
%

\end{document}